\documentclass[12pt]{article}
\usepackage{amsmath}
\usepackage{amsthm}
\usepackage{graphicx,psfrag,epsf}
\usepackage{enumerate}
\usepackage{natbib}
\usepackage{url} 

\newtheorem{remark}{Remark}

\theoremstyle{definition}
\newtheorem{definition}{Definition}[section]
\numberwithin{equation}{section}

\newcommand{\blind}{0}

\newtheorem{condition}{Condition}[section]

\usepackage[T1]{fontenc}
\usepackage[latin9]{inputenc}
\usepackage{geometry}
\usepackage{color}
\usepackage[dvipsnames]{xcolor}
\usepackage{amsthm}
\usepackage{amssymb}
\usepackage{hyperref}

\usepackage{color}

\DeclareMathOperator*{\argmin}{arg\,min}

\begin{document}


\def\spacingset#1{\renewcommand{\baselinestretch}%
{#1}\small\normalsize} \spacingset{1}


\if0\blind
{
  \title{\bf Visually Communicating and Teaching Intuition for
Influence Functions}
  \author{Aaron Fisher 
    \\
    Takeda Pharmaceuticals, \\  Cambridge, MA 02139, (\href{mailto:afishe27@alumni.jh.edu}{afishe27@alumni.jh.edu})\\
    and \\
    Edward H. Kennedy\thanks{
        Authors listed in order of contribution, with largest contribution first.
        At the time of original submission, Aaron Fisher was a postdoctoral fellow in the Department of Biostatistics at the Harvard T.H. Chan School of Public Health. He is currently a statistician at Takeda Pharmaceuticals.
        Edward H. Kennedy is an assistant professor in the Department of Statistics \& Data Science at Carnegie Mellon University.
        Support for this work was provided by the National Institutes of Health (grants 
        P01CA134294, 
        R01GM111339, 
        R01ES024332,
        R35CA197449, 
        R01ES026217,
        P50MD010428,
        DP2MD012722,
        R01MD012769, \&
        R01ES028033), 
        by the Environmental Protection Agency (grants
        83615601 \&
        83587201-0), 
        by the Health Effects Institute (grant
        4953-RFA14-3/16-4),
        and by the National Science Foundation (grant
        DMS-1810979). 
        We are grateful for several helpful conversations with Daniel Nevo, Leah Comment, and Isabel Fulcher, over the course of developing this work.
    } \\
    Department of Statistics \& Data Science at Carnegie Mellon University,\\ Pittsburgh, PA 15213.
    }
  \maketitle
} \fi

\if1\blind
{
  \bigskip
  \bigskip
  \bigskip
  \begin{center}
    {\LARGE\bf Visually Communicating and Teaching Intuition for
Influence Functions}
\end{center}
  \medskip
} \fi

\bigskip
\begin{abstract}
   Estimators based on influence functions (IFs) have been shown to be effective in many settings, especially when combined with machine learning techniques. By focusing on estimating a specific target of interest (e.g., the average effect of a treatment), rather than on estimating the full underlying data generating distribution, IF-based estimators are often able to achieve asymptotically optimal mean-squared error. Still, many researchers find IF-based estimators to be opaque or overly technical, which makes their use less prevalent and their benefits less available. 
   To help foster understanding and trust in IF-based estimators, we present tangible, visual illustrations of when and how IF-based estimators can outperform standard ``plug-in'' estimators. The figures we show are based on connections between IFs, gradients, linear approximations, and Newton-Raphson.
\end{abstract}

\noindent%
\vfill

\newpage
\spacingset{1.45} 

\section{Introduction}

Influence functions (IFs) are a core component of
classic statistical theory, and have emerged as a popular framework
for incorporating machine learning algorithms in inferential tasks \citep{van_der_Laan2011_TMLE_book1,kennedy2017nonparametric, chernozhukov2018double}. 
Estimators based on IFs have been shown to be effective in causal inference and missing data
\citep{robins1994estimation, robins1995semiparametric,van2003unified}, 
regression \citep{van2006statistical_inf_VI,williamson2017nonparametric_VI}, and several other areas \citep{bickel1988estimating,kandasamy2014influence_ML}.

Unfortunately, the technical theory underlying IFs
intimidates many researchers away from the subject. This lack of approachability slows both the theoretical progress within the IF literature, and the dissemination of results.

One typical approach for partially explaining intuition
for IF-based estimators is to describe properties that can be easily
seen in their formulas. For example, IFs can be used to estimate average treatment effects from observational data, after first modeling the process by which individuals are assigned to treatment, and the outcome process that the treatment is thought to affect. The resulting IF-based estimates have been described as ``doubly
robust (DR)'' in the sense that they remain consistent if either the treatment model or the outcome model is correctly specified up to a parametric form \citep{van2003unified, bang2005doubly, kang2007demystifying_DR}. 
While the DR property can sometimes be checked by simply observing an estimator's formula, it does not necessarily provide intuition
for the underlying theory of IF-based estimators. 
Furthermore, the DR property often does not capture an arguably more important benefit of these estimators, which is that they can attain parametric rates of convergence even when constructed based on flexible nonparametric estimators that themselves converge at slower rates. 
Unlike the DR explanation, the notion of faster convergence rates with no parametric assumptions can also extend to applications of IFs beyond the goal of treatment effect estimation \citep{bickel1988estimating,birge1995estimation, kandasamy2014influence_ML, williamson2017nonparametric_VI}.

This paper visually demonstrates a general intuition
for IFs, based on a connection to linear approximations and Newton-Raphson. Our target audiences are statisticians and statistics students who have some familiarity with multivariate calculus. Our hope is that these illustrations can be similarly useful
to illustrations of the standard derivative as the ``slope at a point,''
or illustrations of the integral as the ``area under a curve.'' For
these calculus topics, a guiding intuition can be visualized in minutes,
even though formal study typically takes over a semester of coursework.

In Section \ref{sec:notation} we introduce notation. We also review ``plug-in'' estimators, which will serve as a baseline for comparison. In Sections \ref{sec:Newton's-(1-step)-method} \& \ref{subsec:Sensitivity-to-the}
we show figures illustrating why nonparametric, IF-based estimators
can asymptotically outperform plug-in estimators, but may underperform with small
samples. We avoid heuristic 2-D or 3-D representations of an infinite-dimensional
distribution space, and instead show literal, specific 1-dimensional
paths through that space. In Section \ref{sec:discussion} we briefly discuss connections to semiparametric models, higher order
IFs, and robust statistics. Our overall goal is to facilitate discussion and teaching of
IF-based estimators so that their benefits can be more widely developed
and applied.

\section{Setup: target functionals and ``plug-in'' estimates\label{sec:notation}}

Suppose we observe a sample $z_1,z_2,\dots,z_n$ representing $n$ independent and identically distributed draws of a random vector $Z$ following an unknown distribution $P$. For ease of notation, we will generally assume that $Z$ is continuous, unless otherwise specified in particular examples. We consider the setting where we wish to estimate a particular 1-dimensional ``target'' description of the distribution $P$, also known as an \emph{estimand}. Any such ``target'' can be written as a \emph{functional} of a distribution function, using notation such as $T(P)$. The term ``functional'' 
simply indicates that the input to $T$ is itself a (distribution) function. For example, if $Z=(Z_1,Z_2)$ is bivariate, we may consider the mean of $Z_j$, denoted by $T_{\text{mean},j}(P):=E_P(Z_j)$; the covariance of $Z_1$ and $Z_2$, denoted by $T_{\text{cov}}(P):=E_P(Z_1Z_2)-E_P(Z_1)E_P(Z_2)$; or the conditional expectation of $Z_1$, denoted by $T_{\text{cond},z_2}(P):=E_P(Z_1|Z_2=z_2)$, where $E_P$ is the expectation function with respect to the distribution $P$.

One intuitive approach for estimating functionals $T(P)$ is to simply ``plug-in'' the empirical distribution. This produces the estimate $T(\hat{P})$, where $\hat{P}$ is the distribution placing probability mass $1/n$ at each observed sample point $z_1,\dots,z_n$. While plugging in $\hat{P}$ will suffice for certain estimation targets, such as the mean of a scalar variable $Z$, it is unreliable for other targets, such as the density of a continuous, scalar variable $Z$ at a previously unobserved value $z_{\text{new}}$. The conditional expectation functional described above, $T_{\text{cond},z_2}(P)=E_P(Z_1|Z_2=z_2)$, poses a similar challenge in the bivariate setting. If the value $z_2$ has not been previously observed, then some form of interpolation beyond $\hat{P}$ will be required. Of course, the ``plug-in'' approach easily extends to allow this. Rather than using $\hat{P}$, any smoothed or parametric estimate $\tilde{P}$ of the distribution $P$ can be plugged in to estimate $T(P)$ as $T(\tilde{P})$. Further, if $\tilde{P}$ is a parametric, maximum likelihood estimate (MLE) of $P$, then $T(\tilde{P})$ is a MLE as well, and enjoys similar optimality properties when the likelihood assumptions are correct (by the invariance property of the MLE; see \citealt{casella2002statistical}).

The focus of this paper is on estimation techniques that weaken the likelihood assumptions required for plug-in MLEs. Specifically, we will see that estimates based on influence functions allow us to use flexible estimates for $P$, and to make asymptotic statements about estimator performance, without strict parametric assumptions. Importantly, these IF-based estimates adapt to the particular target of interest $T$, whereas likelihood-based approaches ignore the choice of $T$ (see discussion in Section 1.4 of \citealt{van_der_Laan2011_TMLE_book1}). When likelihood assumptions do not hold, estimators based on influence functions will often converge more quickly than simpler plug-in estimates.

\section{First order based-corrections: visualizing influence functions for estimands \label{sec:Newton's-(1-step)-method}}

Influence functions (IFs) were originally introduced as a description of estimator stability, namely, of how much an estimator changes in response to a slight \emph{perturbation} in the sample distribution (\citealt{hampel1974influence}; see Section \ref{sec:robust}, below).
In the case of plug-in estimators, IFs can also address the parallel, more optimistic question: ``how would the plug-in estimate $T(\tilde{P})$ change in response to a slight \emph{improvement} in our estimate $\tilde{P}$?'' Remarkably, this question can be informed even without directly observing a more accurate version of $\tilde{P}$, as we illustrate in the remainder of this section.

To clarify what we mean by a ``slight improvement'' in $\tilde{P}$, we define a set of distribution estimates indexed by their accuracy. Specifically, let $p$ and $\tilde{p}$ be probability densities for $P$
and $\tilde{P}$ respectively. As in the previous section, $\tilde{P}$ here denotes a smoothed or parametric estimate of $P$. Let  $P_\epsilon$ be the distribution with density
\begin{equation}
p_{\epsilon}(z):=(1-\epsilon)p(z)+\epsilon\tilde{p}(z)\label{eq:P-path-def}
\end{equation}
for $\epsilon \in [0,1]$, where the accuracy of $P_\epsilon$ improves as $\epsilon$ approaches zero. Distributions of this form are sometimes written with the shorthand $P_{\epsilon}:=P+\epsilon(\tilde{P}-P).$ We now refer to the set $\mathcal{P}:=\{P_\epsilon\}_{\epsilon\in[0,1]}$ as a \emph{path} within the space of possible distribution functions that connects $\tilde{P}$ to $P$. For each distribution $P_{\epsilon}$ along this path, there
exists a corresponding value for $T(P_{\epsilon})$, though note that in practice the functional can only be computed at the end point $\epsilon = 1$.

\begin{figure}
\begin{centering}
\includegraphics[width=1\columnwidth]{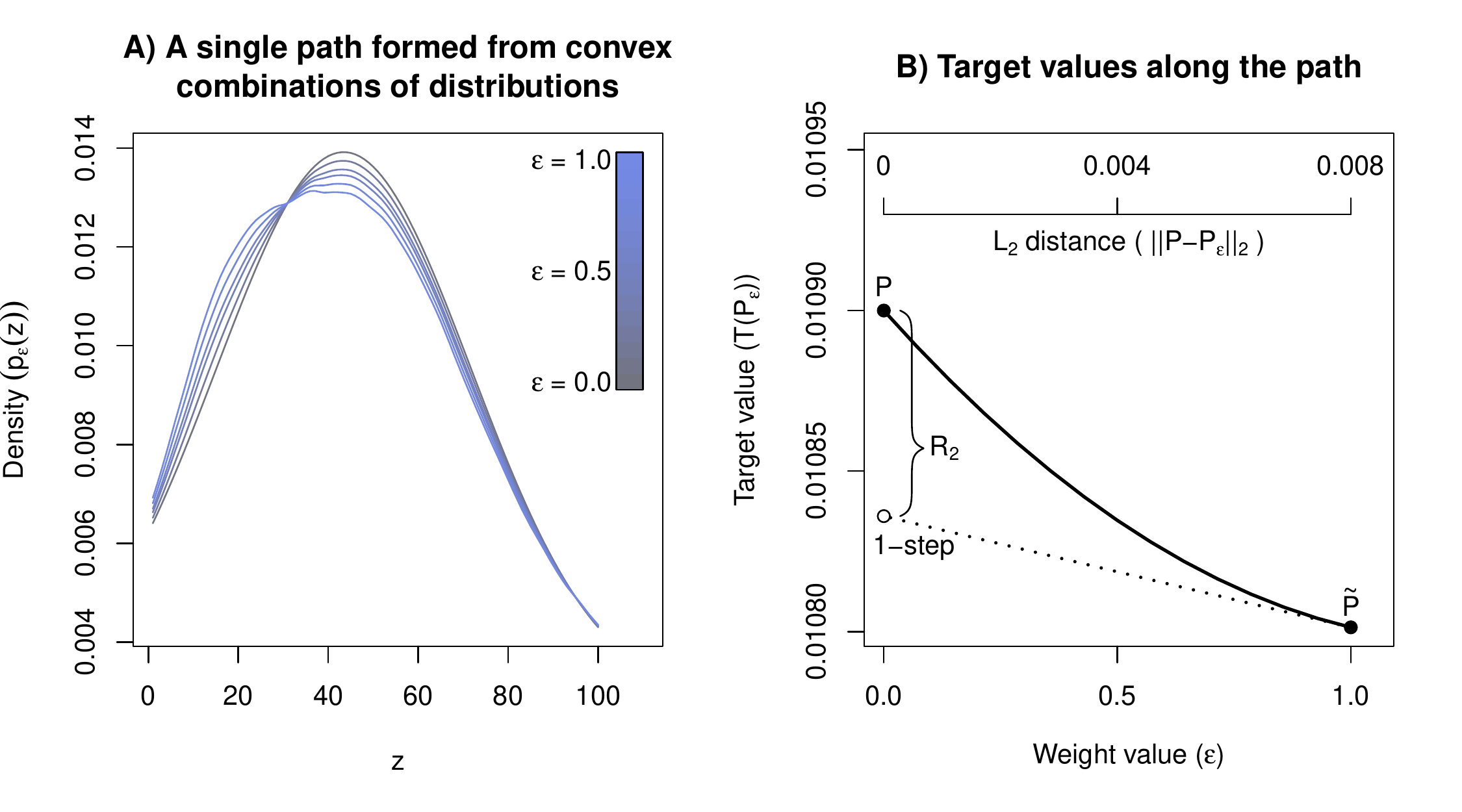}
\par\end{centering}
\caption{Linear approximation of $\mathcal{P}$\label{fig:single-path} -
Given $P$ and $\tilde{P}$, \textbf{Panel A}
shows a subset of the distributions in $\mathcal{P}$ as we vary $\epsilon\in[0,1]$
(see Eq.~(\ref{eq:P-path-def})). When $\epsilon=0$ we have $p_{\epsilon}=p$,
and when $\epsilon=1$ we have $p_{\epsilon}=\tilde{p}$. In \textbf{Panel
B}, the solid line shows the target functional value (y-axis) as we vary $\epsilon$ (x-axis). The dotted line shows the slope of $T(P_{\epsilon})$
with respect to $\epsilon$ at $\epsilon=1$. This slope is calculated using the IF (see Eq.~(\ref{eq:convex-score}), and the Appendix).
Because $||P-P_{\epsilon}||_{2}=\epsilon||P-\tilde{P}||_{2}$
(see Section \ref{subsec:Sensitivity-to-the}, and the Appendix), the x-axis can equivalently be
expressed either in terms of $||P-P_{\epsilon}||_{2}$
or in terms of $\epsilon$. Reflecting this, we show the distributional distance
$||P-P_{\epsilon}||_{2}$ on a secondary horizontal axis at the top
of the figure. 
}
\end{figure}

We illustrate an example of such a set of distributions
in Figure \ref{fig:single-path}-A, and illustrate the values $T(P_{\epsilon})$
along this path in Figure \ref{fig:single-path}-B. As a working example
for our illustrations, we will use the functional of the integrated
squared density, $T(P)=\int p(z)^{2}dz$, for a 1-dimensional variable $Z$  \citep{bickel1988estimating,birge1995estimation,laurent1996efficient,gine2008simple,robins2009quadratic}. This is purely
for the purposes of coding an example figure however.
The technical discussion below does not assume $T(P)=\int p(z)^{2}dz$. In the appendix, we additionally illustrate the special case where $Z$ is discrete, and where we can show the space of all possible distributions in a 2-dimensional figure.

Our ultimate goal is to find the y-intercept of the curved,
solid line in Figure \ref{fig:single-path}-B. We denote this line by the function $v$, where $v(\epsilon):=T(P_\epsilon)$ and the y-intercept of interest is $v(0)=T(P_{0})=T(P)$. 
Fortunately, although the solid curve $v(\epsilon)$ is unknown and can only be evaluated
at $\epsilon=1$, we will see shortly that it is still possible to \emph{approximate} this curve, and to find the y-intercept of our approximation. Specifically, we will see that we can estimate the slope of $v(\epsilon)$
 at $\epsilon=1$, denoted here by $v'(1):=\left.\frac{\partial}{\partial\epsilon}T(P_{\epsilon})\right|_{\epsilon=1}$. This, in turn, lets us approximate the curve $v(\epsilon)$
linearly at $\epsilon=1$. The y-intercept for our approximation of $v$
is then equal to $T(P_{1})-dv'(1)$ (shown as
``1-step'' in Figure \ref{fig:single-path}-B),
where $d=1$ is the distance between $P_{1}$ and $P_{0}$
in terms of $\epsilon$. Thus, an ideal estimator for $T(P_0)$ might resemble $\{T(P_{1})-v'(1)\}$, motivated by how our plug-in estimate ($T(P_1)$) would change if our initial distribution estimate ($P_1$) became infinitesimally more accurate ($-v'(1)$). Before considering how $v'(1)$ may be estimated, we discuss two interpretations of this ``1-step'' approach (see also \citealt{bickel_1975_one_step,kraft1972asymptotic} for early examples of 1-step estimators).

One understanding of the ``1-step'' approach comes from an analogy to Newton-Raphson -- an iterative procedure for finding the roots of a real function $f$. Given an initial guess $x_0\in\mathbb{R}$ of a root of $f$ (a value $x_{\text{root}}$ satisfying $f(x_{\text{root}})=0$), Newton-Raphson attempts to improve on this guess by approximating $f$ linearly at $x_0$. The root of this linear approximation is taken as an updated guess for a root of $f$, and the procedure is iterated until convergence.
When $v$ (defined above) is invertible, finding the value of $T(P)=v(0)$
is equivalent to a root-finding problem for $v^{-1}$, and the ``1-step''
method described above is equivalent to 1 step of Newton-Raphson for the function
$v^{-1}$ (see \citealp{pfanzagl1982contributions}).

The ``1-step'' approach
can also be motivated from the Taylor expansion of the function $v$:
\begin{align}
T(P_{0})=v(0) & =v(1)+v'(1)(0-1)-R_{2}\nonumber \\
 & =T(P_{1})+\left.\frac{\partial}{\partial\epsilon}T(P_{\epsilon})\right|_{\epsilon=1}(0-1)-R_{2},\label{eq:1dim-taylor-eq}
\end{align}
where $R_{2}=(-1/2)v''(\bar{\epsilon})=(-1/2)\left.\frac{\partial^{2}}{\partial\epsilon^{2}}T(P_{\epsilon})\right|_{\epsilon=\bar{\epsilon}}$
for some value $\bar{\epsilon}\in[0,1]$ by Taylor's theorem \citep{serfling1980approximation}.\footnote{Absorbing a negative sign into the definition of $R_2$ will help to simplify residual terms later on.} The first two terms in Eq.~(\ref{eq:1dim-taylor-eq}) are equal to $T(\tilde{P})-v'(1)$, reproducing the ``1-step approach'' described above, and the remaining $R_{2}$ term can typically be shown to be small. Formally studying $R_{2}$ via 
Taylor's Theorem requires that
$v'$ and $v''$ are finite, and that $v'$ is continuous, although
these conditions are not necessary if the $R_{2}$ term can
instead be studied directly (see Section \ref{subsec:Sensitivity-to-the};
and \citealp{serfling1980approximation}). 
Because our 1-step approach $T(P_{1})-v'(1)$ uses only on the first derivative of $v(\epsilon)=T(P_{\epsilon})$, we refer to it as a \emph{first order bias-correction}. We refer to this derivative as a \emph{pathwise derivative} along $\mathcal{P}$. We now turn to the task of estimating this derivative, which is precisely where IFs will prove useful.

We start with the case when $Z$ is a discrete random variable, as this makes estimation of $v'(1)=\left.\frac{\partial}{\partial\epsilon}T(P_{\epsilon})\right|_{\epsilon=1}$ appear relatively straightforward.
Let $\{z_1,\dots,z_K\}$ be the set of values that $Z$ may take. With some abuse of notation, we can determine the derivative $\left.\frac{\partial}{\partial\epsilon}T(P_{\epsilon})\right|_{\epsilon=1}$ from the partial derivatives of $T(P_{\epsilon})$ with respect to each value of the probability mass function $p_{\epsilon}(z_k)$, using the multivariate chain rule:
\begin{align}
\left.\frac{\partial}{\partial\epsilon}T(P_{\epsilon})\right|_{\epsilon=1} 
& =\sum_{k=1}^K \frac{\partial T(P_{\epsilon})}{\partial p_{\epsilon}(z_k)}\left.\frac{\partial p_{\epsilon}(z_k)}{\partial\epsilon}\right|_{\epsilon=1} \label{eq:chain-rule}\\
 &=\sum_{k=1}^K \left.\frac{\partial T(P_{\epsilon})}{\partial p_{\epsilon}(z_k)}\right|_{\epsilon=1}
\{\tilde{p}(z_k)-p(z_k)\}.
\label{eq:if-grad}
\end{align}
Eq.~(\ref{eq:chain-rule}) states that the change in $T(P_\epsilon)$ depends on how $T(P_\epsilon)$ changes with each probability mass $p_\epsilon(z_k)$, and on how each probability mass changes with $\epsilon$. 
However, the above equation is an abuse of notation in the sense that marginal increases to $p_\epsilon(z_k)$ result in $p_\epsilon$ no longer being a valid probability mass function (its total mass will not equal 1), which can cause the partial derivatives $\frac{\partial T(P_{\epsilon})}{\partial p_{\epsilon}(z_k)}$ to be ill-defined. Any marginal additional mass at $p(z_k)$ must instead be accompanied by an equal decrease in mass elsewhere in the distribution.

This shortcoming of the partial derivatives of $T$ motivates us to replace them with the  \emph{influence function} for $T$, defined below (see \citealt{kandasamy2014influence_ML}, and Section 6.3.1 of \citealt{serfling1980approximation}).

\begin{definition}\label{def:if-diff-def}
For a given functional $T$, the \emph{influence function} for $T$ is the function $IF$ satisfying
\begin{align}
\left. \frac{\partial T(G + \epsilon(Q-G))}{\partial \epsilon} \right|_{\epsilon=0} = 
\int IF(z,G) \{q(z)-g(z)\}dz \label{eq:if-diff-def}
\end{align}
and $\int IF(z,G)g(z)dz=0$ 
for any two distributions $G$ and $Q$ with densities $g$ and $q$. Above, $G + \epsilon(Q-G)$ denotes the distribution with density function $g(z)+\epsilon(q(z)-g(z))$, as defined in Eq.~(\ref{eq:P-path-def}).
\end{definition}
Roughly speaking, the left-hand side of Eq.~(\ref{eq:if-diff-def}) is the change in $T(G)$ that would occur if we were to ``mix'' $G$ with an infinitesimal portion of the distribution $Q$. This quantity is known as the \emph{G\^{a}teaux derivative} \citep{serfling1980approximation}, and can be interpreted as the sensitivity of $T(G)$ to small changes in the underlying distribution $G$, in the ``direction'' of $Q$.

The IF in Eq.~(\ref{eq:if-diff-def}) has a similar interpretation to the partial derivative in Eq.~(\ref{eq:if-grad}). To see this, we can isolate the IF term $IF(z,G)$ by setting $Q$ equal to the point mass distribution at $z$, denoted by $\delta_z$ (see \citealt{hampel1974influence,van_der_Vaart2000asymptotic_statistics}). 
Here, Eq.~(\ref{eq:if-grad}) reduces to 
\begin{align}
\left. \frac{\partial T(G + \epsilon(\delta_z-G))}{\partial \epsilon} \right|_{\epsilon=0} =  IF(z,G).\label{eq:IF-single-z}
\end{align} 
The left-hand side is the change in $T(G)$ that would occur in response to infinitesimally upweighting $z$, analogous to the interpretation of the partial derivative in Eq.~(\ref{eq:if-grad}) (see also Section 6.3.1 of \citealt{serfling1980approximation}). With this analogy in mind, note the similarity between the right-hand sides of Eq.~(\ref{eq:if-grad}) and Eq.~(\ref{eq:if-diff-def}). Roughly speaking, the IF lets us apply the ``multivariate chain rule'' approach from Eq.~(\ref{eq:if-grad}), but remains well defined even when the partial derivatives in Eq.~(\ref{eq:if-grad}) are not. 

A common alternative (though in many cases equivalent) ``score-based'' definition of the IF is presented in the Appendix (see \citealt{bickel1993efficient,tsiatis2006semiparametric}). This definition allows the IF to directly describe derivatives along more general pathways of distributions, extending beyond pathways of the form $G+\epsilon(Q-G)$. Such pathways become of particular interest in cases where prior knowledge restricts the space of distributions that we consider possible, and where this restricted space is not closed under mixture of distributions (see discussion in Section \ref{sec:semipara}).

Returning to our example of the pathway $\mathcal{P}$, we
can now use the IF to derive
an empirical estimate of $\left.\frac{\partial}{\partial\epsilon}T(P_{\epsilon})\right|_{\epsilon=1}$ (e.g., the dashed line in Figure \ref{fig:single-path}). Applying Eq.~(\ref{eq:if-diff-def}), we have\footnote{To apply Eq.~(\ref{eq:if-diff-def}) in Eq.~(\ref{eq:convex-score}), we rearrange $\left.\frac{\partial}{\partial\epsilon}T(P_{\epsilon})\right|_{\epsilon=1}$  as 
$\left.\frac{\partial}{\partial\epsilon}T(P+\epsilon(\tilde{P}-P))\right|_{\epsilon=1}=-
\left.\frac{\partial}{\partial a}T(\tilde{P}+a(P-\tilde{P}))\right|_{a=0}$.}  
\begin{align}
\left.\frac{\partial}{\partial\epsilon}T(P_{\epsilon})\right|_{\epsilon=1} 
&=-\int IF(z,\tilde{P})\left\{ p(z)-\tilde{p}(z)\right\} dz  &  & \label{eq:convex-score} \\
 & =-\int IF(z,\tilde{P})p(z)dz &  & \text{from }\int IF(z,\tilde{P})\tilde{p}(z)dz =0\nonumber \\
 & \approx-\frac{1}{n}\sum_{i=1}^{n}IF(z_{i},\tilde{P}).\label{eq:IF-approx-sample}
\end{align}

In this way, IFs can provide estimates (Eq.~(\ref{eq:IF-approx-sample})) of distributional derivatives (Eq.~(\ref{eq:convex-score}), which corresponds to the dashed line in Figure \ref{fig:single-path}). Studying these estimates is fairly straightforward if $\tilde{P}$ can be treated as fixed, for instance, if $\tilde{P}$ is estimated a priori or using sample splitting. In such cases, we can treat Eq.~(\ref{eq:IF-approx-sample}) as a simple sample average. Alternatively, if we allow
the current dataset $\{z_{1},\dots,z_{n}\}$ to inform the selection
of $\tilde{P}$ as well as the calculation of the summation
in Eq.~(\ref{eq:IF-approx-sample}), then formal study of the estimator in Eq.~(\ref{eq:IF-approx-sample}) is still possible as long as $\tilde{P}$
is selected from a sufficiently regularized class (e.g., a Donsker
class). In this case, the bias and variance of $\sum_{i=1}^{n}IF(z_{i},\tilde{P})$
can be studied using empirical process theory \citep{van_der_Vaart2000asymptotic_statistics}. Hereafter, we assume the simpler case where $\tilde{P}$ is estimated a priori, and can be treated as fixed.

Combining the results from Eq.~(\ref{eq:1dim-taylor-eq})
and \ref{eq:IF-approx-sample}, we can approximate $T(P)$
using our dataset, as
\begin{equation}
T(P)\approx T(\tilde{P})+\frac{1}{n}\sum_{i=1}^{n}IF(z_{i},\tilde{P})-R_{2},\label{eq:approx-T0-avg}
\end{equation}
where the approximation symbol captures the fact that we are using a sample average. This motivates the ``1-step'' estimator 
\[
\hat{T}_{\text{1-step}}:=T(\tilde{P})+\frac{1}{n}\sum_{i=1}^{n}IF(z_{i},\tilde{P}).
\]
Conditions under which the $R_{2}$ term converges to zero are discussed
in the next section. 

We can see from Figure \ref{fig:single-path} that
when $R_{2}$ is in fact negligible, the only challenge remaining
is to estimate the slope $\left.\frac{\partial}{\partial\epsilon}T(P_{\epsilon})\right|_{\epsilon=1}$,
which can be done in an unbiased and efficient way via Eq.~(\ref{eq:IF-approx-sample}).
It should not be surprising then that the estimator $\hat{T}_{\text{1-step}}$,
which takes precisely this approach, has optimal mean-squared error
(MSE) properties when $R_{2}$ is small. More specifically, given
no parametric assumptions on $P$, it can be shown that \emph{no
estimator of} $T(P)$ can have a MSE
uniformly lower than $n^{-1}\text{Var}(IF(z,P))$. We refer to \citet{van_der_Vaart2000asymptotic_statistics,van2002part} for more details on this minimax lower bound result. In practice, the variance
bound $n^{-1}\text{Var}(IF(z,P))$ can be approximated by $n^{-1}\text{Var}(IF(z,\tilde{P}))=\text{Var}(\hat{T}_{\text{1-step}})$.
Thus, when $R_{2}$ is negligible and $\text{Var}(IF(z,\tilde{P}))$
approximates $\text{Var}(IF(z,P))$ well, estimating the
slope through $\tilde{P}$ yields an approximately unbiased and efficient estimator. 

\section[Visualizing the residual, and the sensitivity
to the choice of initial distribution estimator]{Visualizing the residual $R_{2}$, and the sensitivity
to the choice of initial estimator $\tilde{P}$\label{subsec:Sensitivity-to-the}}

Formal study of the $R_{2}$ term is often done
on a case-by-case basis by algebraically simplifying the residual
$E_{P}\left\{\hat{T}_{\text{1-step}} - T(P)\right\} $, and so
Taylor's Theorem is often not needed to describe the $R_{2}$ term
(Eq.~(\ref{eq:1dim-taylor-eq})). In many cases, the $R_{2}$ term reveals itself to be a quadratic combination of one or more error terms. For example, for the integrated squared
density functional $T(P)=\int p(z)^{2}dz$, the $R_{2}$
term can be shown to be exactly equal to the negative of $\int\{p(z)-\tilde{p}(z)\}^{2}dz$
(see the Appendix). When the error term $p(z)-\tilde{p}(z)$ converges (uniformly) to zero, the $2^{nd}$ degree exponent implies that $R_2$ converges to zero even more quickly.

\begin{figure}
\begin{centering}
\includegraphics[width=1\columnwidth]{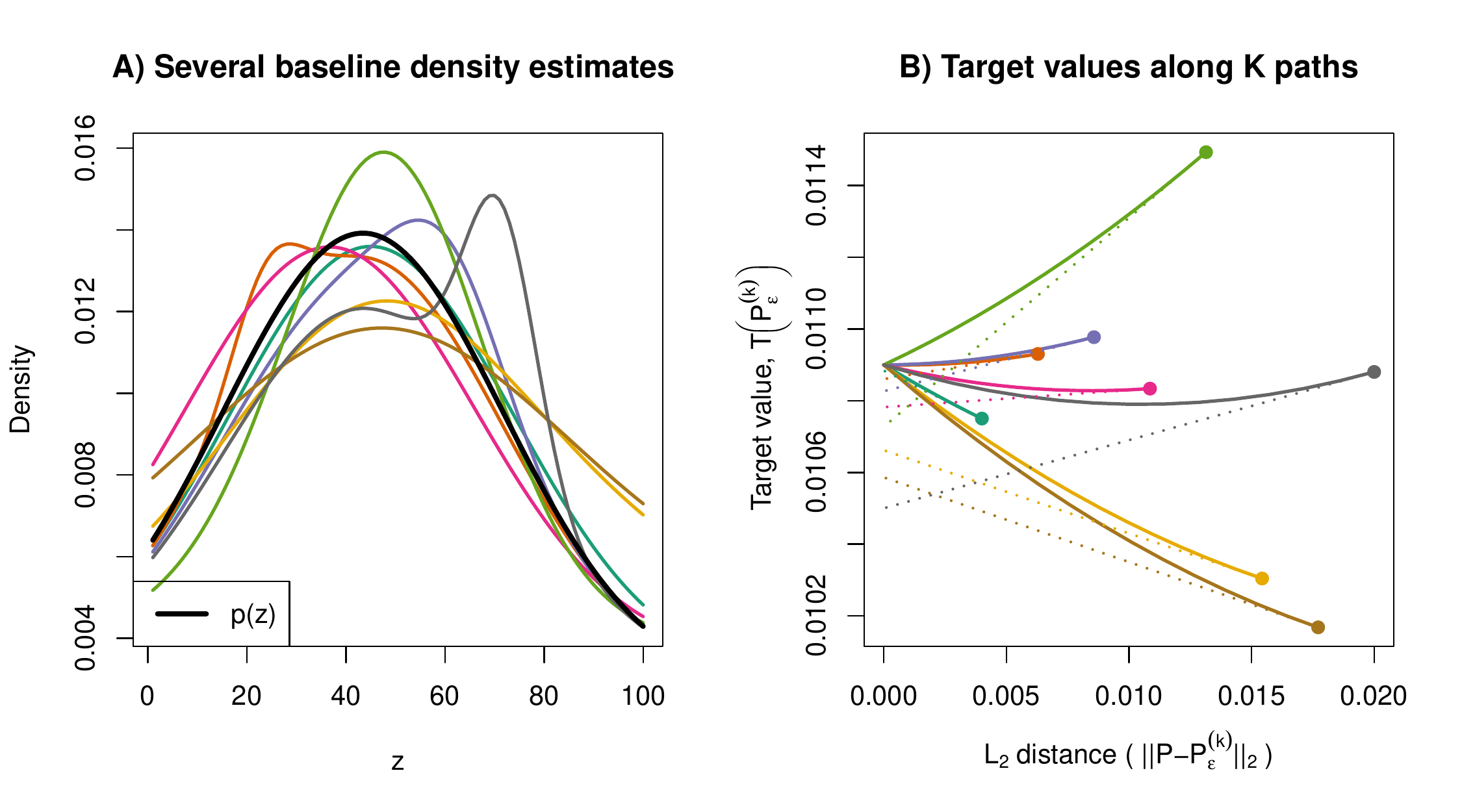}
\par\end{centering}
\caption{Linear approximations overlaid for several paths \label{fig:several-paths}
- \textbf{Panel A} overlays the same illustration as Figure \ref{fig:single-path}-A,
but for several alternative initial distribution estimates $\tilde{P}^{(1)},\dots,\tilde{P}^{(K)}$.
For each distribution $\tilde{P}^{(k)}$, a path $\mathcal{P}^{(k)}$
connecting $P$ to $P^{(k)}$ can be defined in
the same way as $\mathcal{P}$. \textbf{Panel B} shows the values
of the target parameter at each point $\tilde{P}_{\epsilon}^{(k)}$
along each path $\mathcal{P}^{(k)}$, as well as a linear approximation
of each path. For each value of $k\in 1,\dots,K$, we show the distribution $\tilde{P}^{(k)}$ (Panel A) and pathway $\mathcal{P}^{(k)}$ (Panel B) in the same color. On the x-axis in Panel B, we plot each distribution's distance
from $P$, in order to show several paths simultaneously. The y-intercept of each linear approximation corresponds to a different
1-step estimator, and the accuracy of this estimator will depend on
the distance $||P-\tilde{P}^{(k)}||_{2}$.}
\end{figure}

A similar result can be shown for the general case of smooth functionals $T$. Here, $R_{2}$ will turn out to depend
on two pieces of information that make the problem difficult: the
underlying distributional distance between $\tilde{P}$ and
$P$, which is typically assumed to converge to zero as sample
size grows, and the ``smoothness'' of $T$ (defined below). In
the remainder of this section we visually illustrate this result
(Figure \ref{fig:several-paths}), and review this result formally.

Figure \ref{fig:several-paths} shows how Figure
\ref{fig:single-path} would change if we had selected an initial
distribution estimate different from $\tilde{P}$. Figure
\ref{fig:several-paths}-A shows several alternative distribution
estimates, denoted by $\tilde{P}^{(k)}$ for $k=1,\dots,K$.
For each initial estimate $\tilde{P}^{(k)}$, we define the
path $\mathcal{P}^{(k)}$ as the set of distributions $P_{\epsilon}^{(k)}=(1-\epsilon)P+\epsilon\tilde{P}^{(k)}$
for $\epsilon\in[0,1]$, analogous to $\mathcal{P}$. Figure \ref{fig:several-paths}-B
shows each of these $K$ paths, as well as the 1-step estimators corresponding
to each path. We can see that the 1-step estimators are generally
more effective when $\tilde{P}^{(k)}$ is ``closer'' to
$P$ (defined formally below). We can also see that, as in Figure \ref{fig:single-path}, the performance of 1-step estimators depends on the smoothness of $T(P_{\epsilon}^{(k)})$ with respect to $\epsilon$.

Quite informally, we can think of Figure \ref{fig:several-paths}-B as a ``Magician's Tablecloth Pull-Plot.'' To see this analogy, try to imagine the functional $T$ as a hyper-surface over the space of possible distributions. (In the Appendix, we illustrate a special case where this hyper-surface reduces to a standard 3-dimensional surface.) Then, imagine a magician pinching this surface at the point $P$, and pulling the surface to one side as one might dramatically pull a tablecloth from a table, with the unpinched fabric folding in on itself as it billows in the air. As we watch this pulling action (e.g., from a neighboring table), all of the dimensionality of the hyper-surface folds into 1 dimension: how far each point on the surface (or ``fabric'') is from the distribution $P$ (the point the magician is pulling from). In Figure \ref{fig:several-paths}-B, we can imagine the intersection point on the left-hand side as the point from which the magician is pulling the tablecloth.

To formalize the notion of how ``far'' two distributions
$G$ and $Q$ are, we use the $L_2$ distance $||G-Q||_{2}:=\sqrt{\int[g(z)-q(z)]^{2}dz}$,
where $g$ and $q$ are the densities of $G$ and $Q$
respectively. 

This distance measure is useful in part because it lets us visually overlay several paths with a common, meaningful x-axis (Figure \ref{fig:several-paths}), and in part because it helps us formally compare the ``smoothness'' of $T$ along paths that stretch over different distances. Recall that the path $\{P_\epsilon\}_{\epsilon\in[0,1]}$ connects the two distributions $\tilde{P}$ and $P$, which are a distance of $||\tilde{P}-P||_2$ from each other. One approach for describing the smoothness of $T$ is to consider how quickly $T(P_\epsilon)$ changes in response to changes in $\epsilon$, but this notion of smoothness is highly sensitive to our choice of $\tilde{P}$ -- the starting point of our pathway. For example, if we were to move $\tilde{P}$ closer to $P$, then $T$ would appear to be smoother. In order to describe the smoothness of $T$ in a way that is not sensitive to the choice of $\tilde{P}$, we consider the following reindexing of $P_\epsilon$. Let
\begin{equation}
P_{\Delta}^\text{rescaled}:=P+\left(\frac{\Delta}{||\tilde{P}-P||_{2}}\right)(\tilde{P}-P),\hspace{.6cm}\text{for } \Delta\in[0,||\tilde{P}-P||_{2}].\label{eq:rescaled-intro} 
\end{equation}
This definition produces the same pathway as in Eq.~(\ref{eq:P-path-def}), as $P_{\epsilon}=P_{\Delta}^{\text{rescaled}}$ when $\epsilon=\Delta/||\tilde{P}-P||_{2}$. However, it can be shown that $\Delta$ tells us the absolute distance $\Delta=||P_{\Delta}^\text{rescaled}-P||_2$, whereas $\epsilon$ tells us the relative distance $\epsilon=||P_{\epsilon}-P||_2/||\tilde{P}-P||_2$ (see the Appendix). In this way, the information represented by $\Delta$ is less dependent on the choice of $\tilde{P}.$ 

We can now describe the smoothness of $T$ more formally, using the following condition on its $j^{th}$ derivative with respect to the distance-adjusted parameter $\Delta$.
\begin{condition}\label{cond:no-squiggles}
($j^{th}$ order smoothness from all directions) For a given value of $j$, and for any choice of $\tilde{P}$, the function $T\left(P_{\Delta}^\text{rescaled}\right)$ is $j$-times differentiable with respect to $\Delta$, and 
 $\left.\frac{\partial^{j}}{\partial \Delta^{j}}T\left(P_{\Delta}^\text{rescaled}\right)\right|_{\Delta=\bar{\Delta}}
= O(1)$
as $\bar{\Delta}\rightarrow0$.
\end{condition}

For $j=1$, Condition \ref{cond:no-squiggles} bounds the degree to which $T(P)$ can change in response to any small change to $P$. In Figure \ref{fig:several-paths}, this means that curves cannot deviate too far from flat lines as they approach the leftmost region. 
For $j=2$, Condition \ref{cond:no-squiggles} bounds the degree to which $T(P)$ can change \emph{nonlinearly} in response to any small change in $P$. That is, curves cannot get ``too squiggly'' as they approach the leftmost region of Figure \ref{fig:several-paths}.  Note that, for notational convenience, have suppressed the dependence of $P_{\Delta}^\text{rescaled}$ on $\tilde{P}$ in Eq.~(\ref{eq:rescaled-intro}) \& Condition \ref{cond:no-squiggles}. 

The connection between Condition \ref{cond:no-squiggles} and estimator performance can be formalized as follows.
\begin{remark}\label{rem:R1R2}
(Asymptotic bias of plug-in and 1-step estimators) If $\tilde{P}$ is fixed in advance (for example, from sample splitting), and if Condition \ref{cond:no-squiggles} holds for $j=2$, then the bias for $\hat{T}_{\text{1-step}}$ is equal to
\begin{equation}
R_2 = E_{P}(\hat{T}_{\text{1-step}}) - T(P) = O(||P-\tilde{P}||_{2}^{2}).\label{eq:remark-R2}
\end{equation}
Similarly, if $\tilde{P}$ is fixed and Condition \ref{cond:no-squiggles} holds for $j=1$, then the error of the plug-in estimate is equal to
\begin{equation}
T(\tilde{P}) - T(P)  = O(||P-\tilde{P}||_{2}).\label{eq:remark-R1}
\end{equation}
Since we treat $T(\tilde{P})$ as fixed, given $\tilde{P}$, the error of $T(\tilde{P})$ (Eq.~(\ref{eq:remark-R1})) is also equal to the bias of $T(\tilde{P})$.
\end{remark}
To explain in words, as $\tilde{P}$ approaches $P$, the biases of plug-in estimators and 1-step estimators are both guaranteed to converge to zero. However, the worst-case rate of convergence for 1-step estimators is substantially faster than that of plug-in estimators ($O(||P-\tilde{P}||_{2}^2)$ relative to $O(||P-\tilde{P}||_{2})$). The proof of Remark \ref{rem:R1R2} follows from Taylor's Theorem (see the Appendix, as well as Eq.~(1) of \citealt{robins2008higher} for a similar discussion).

Results similar to Remark \ref{rem:R1R2} are often expressed by instead defining the influence function as the unique function $IF$ satisfying
\begin{equation}
T(\tilde{P}) - T(P)  
=  
\int IF(z,\tilde{P}) \ d(\tilde{P}(z)-P(z))
+ R_2(\tilde{P},P),\label{eq:dist-taylor}
\end{equation}
and $E_{P}[IF(z,P)]=0$ for any two distributions $\tilde{P},P$, where $R_2$ satisfies either  $R_2(\tilde{P},P)=O(||P-\tilde{P}||_2^2)$ or a similar condition. Eq.~(\ref{eq:dist-taylor}) is often referred to either as the \emph{distributional Taylor expansion} of $T$, or as the \emph{von Mises expansion} of $T$ \citep{mises1947asymptotic,serfling1980approximation,robins2008higher,robins2009quadratic,fernholz1983mises_book_older,carone2014higher,robins2017minimax}. The expansion is analogous to the standard Taylor expansion in Eq.~(\ref{eq:1dim-taylor-eq}), but plugs in the integral term from Eq.~(\ref{eq:convex-score}).

To obtain a more complete view of 1-step estimators, we must consider the convergence rate of $R_2$ \emph{in combination} with the convergence rate of the sample average in Eq.~(\ref{eq:approx-T0-avg}). Whichever of these two rates is slower will determine the asymptotic behavior of the 1-step estimator.  To see why, recall from Eq.~(\ref{eq:approx-T0-avg}) that the error of the 1-step estimator is equal to
\begin{align} 
\hat{T}_{\text{1-step}} - T(P) 
&=  \left[ \frac{1}{n} \sum_{i=1}^{n} IF(z_i;\tilde{P})   -  E_{P} IF(Z;\tilde{P}) \right] + R_2,\label{eq:expansion-error-simple}
\end{align}
where the bracketed term is a centered sample average that is asymptotically normal after $\sqrt{n}$ scaling. Here we have implicitly assumes that sample splitting has been used to estimate $\tilde{P}$; if not, then the bracketed term can be rearranged and studied using empirical process theory.\footnote{
To account for estimation of $\tilde{P}$, the bracketed term in Eq.~(\ref{eq:expansion-error-simple}) can be written as
\begin{align*} 
\frac{1}{n} \sum_{i=1}^{n} \left[ 
\{IF(z_i,\tilde{P})  - IF(z_i,P)\}
 - E_{P} \{ IF(Z,\tilde{P})  - IF(Z,P)\}\right] 
+ 
\frac{1}{n} \sum_{i=1}^{n} \left[ 
IF(z_i,P) - E_P(Z,P)
\right], 
\end{align*}
Note that both summations are centered around their expectation. The first summation can be studied using empirical process theory, and the second summation can be studied as a simple sample average (see, for example,  \citealt{van2006targeted,van_der_Vaart2000asymptotic_statistics}).
} 
The $R_2$ term is the second-order remainder described in Eq.~(\ref{eq:1dim-taylor-eq}) and (\ref{eq:remark-R2}), which depends on the smoothness of $T$ and the accuracy of $\tilde{P}$. Finite-sample bounds (e.g., using concentration inequalities on the bracketed term, and functional-specific bounds on $R_2$) could be used to construct confidence intervals valid for any $n$. However this would require precise knowledge of the error in $\tilde{P}$ as well as bounds on or variance of the IF, and such intervals may be quite wide in realistic examples. The most common approach in practice is therefore to assume the $R_2$ term (and any empirical process terms) are negligible, and assume the bracketed term in Eq.~(\ref{eq:expansion-error-simple}) can be well-approximated by a normal distribution with appropriate variance. If $R_2$ = $o_P(1/\sqrt{n}$) then this will often be a reasonable approximation at least with large sample sizes, where the specific meaning of ``large'' could be assessed via simulations. However, if $R_2 = O_P(1/n^\alpha)$ for some $\alpha < 1/2$, then such an approximation will not even be asymptotically valid -- the first-order correction is not enough in this case, and instead either sensitivity analyses or higher-order corrections are required (see Section \ref{sec:semipara}, and \citealt{robins2008higher,robins2009quadratic,carone2014higher,robins2017minimax}). 

In summary, the performance of 1-step estimators depends on the sample size (via the sample average in Eq.~(\ref{eq:IF-approx-sample})), the smoothness of the functional of interest ($T$), and the quality of the initial distribution estimate ($\tilde{P}$). Graphically, we can visualize the smoothness of $T$ by the bumpiness of the paths shown in Figure \ref{fig:several-paths}-B. We show the quality of the initial distribution estimate ($\tilde{P}$) by the x-axis in Figure \ref{fig:several-paths}-B.\footnote{Also see Figure \ref{fig:several-3d-paths} in the Appendix.} Reasonably accurate estimates of $\tilde{P}$ land us in the leftmost region of Figure \ref{fig:several-paths}-B, where bias corrections are especially effective. Inaccurate initial estimates, i.e., slow convergence rates due to high-dimensionality, land us in the rightmost area of Figure \ref{fig:several-paths}-B, where linear corrections based on IFs are least effective.

\section{Discussion\label{sec:discussion}}

In this section we briefly review extensions and other uses of IFs. For deeper treatments of
IFs and related topics, interested readers can see \citep{serfling1980approximation,pfanzagl1982contributions, bickel1993efficient,van_der_Vaart2000asymptotic_statistics,van2003unified,tsiatis2006semiparametric,huber2009robust_statistics_2nd_ed_book,kennedy2016semiparametric,maronna2019robust_stat_book}.

\subsection{Semiparametric models\label{sec:semipara}}

Thus far, we have considered so-called \emph{nonparametric} models, in which no a priori knowledge or restrictions
are assumed about the distribution $P$. In certain
cases though, we may already know certain parameters of the probability
distribution. For example, we may know the process by which patients
are assigned to different treatments in a particular cohort, but may not 
know the distribution of health outcomes under each treatment. This
more general framework is known as a  \emph{semiparametric}
model, with the nonparametric model forming a special case of no
priori knowledge.

When some parameters of $P$ are known, the distributions along the path $\mathcal{P}$ may not all satisfy the restrictions enforced by that knowledge. 
We can encode these restrictions in the form of a likelihood assumption, and focus our attention only on pathways of distributions concordant with this likelihood. 
Because we only need to consider derivatives along allowed pathways, the function $IF$ no longer needs to be valid for all distributions $G$ and $Q$ (see Definition \ref{def:if-diff-def}), and can instead be defined in terms of the score function for the likelihood (see the Appendix). 
This relaxed criteria for the influence function will now be met not just by a single function $IF$, but by a \emph{set} ($\mathcal{S}$)
of functions. Of these, if we can identify the ``efficient influence
function'' $IF^{\star}$ equal to $\argmin_{\widetilde{IF}\in\mathcal{S}}\text{Var}(\widetilde{IF}(Z,P))$,
then we can more efficiently estimate the derivatives along allowed
pathways. We can also show that no unbiased estimator may have a variance
lower than $n^{-1}\text{Var}(IF^{\star}(Z,P))$, which is
equal to or lower than the nonparametric bound described above ($n^{-1}\text{Var}(IF(z,P))$).
Determining $IF^{\star}$ requires a projection operation that
is usually the focus of figures illustrating the theory of influence functions (see Sections 2.3 \& 3.4 of \citealp{tsiatis2006semiparametric}), but
this operation is beyond the scope of this paper.

\subsection{Higher order influence functions\label{subsec:HO-IFs}}

The approach of Section \ref{sec:Newton's-(1-step)-method}
amounts to approximating $T(P_{\epsilon})$ as a \emph{linear}
function of $\epsilon$, but several alternative approximations of
$T(P_{\epsilon})$ exist as well. For example, the standard
``plug-in'' estimator $T(\tilde{P})$ can be thought
of as approximating $T(P_{\epsilon})$ as a \emph{constant}
function of $\epsilon$, and extrapolating this approximation to estimate
$T(P_{0})$. Given that the linear approximation often
gives improved estimates over the constant approximation, we might
expect that a more sophisticated approximation $T(P_{\epsilon})$
would improve accuracy even further. Indeed, for the special case
of the squared density functional $T(P)=\int p(z)^{2}dz$
shown in Figures \ref{fig:single-path} \& \ref{fig:several-paths},
a second degree polynomial approximation of $T(P_{\epsilon})$
fully recovers the original function with no approximation error.
In general, deriving higher order polynomial approximations requires
that we are able to calculate higher order derivatives of $T(P_{\epsilon})$,
which forms part of the motivation for recent work on higher order
influence functions. 

Interestingly, it turns out that using higher-order influence functions is not as straightforward as the first-order case, simply because higher-order influence functions do not exist for most functionals of interest (e.g., the integrated density squared, average treatment effect, etc.). In other words, although there is often a function $IF$ satisfying
$$
T(\tilde{P}) - T(P)  = 
\int IF(z,\tilde{P}) \ d(\tilde{P}(z)-P(z)) 
+ R_2(\tilde{P},P),
$$
for an appropriate second-order term $R_2(\tilde{P},P)$ (though not always - see for example \citet{kennedy2017nonparametric}), there is typically no function $IF_{2}$ satisfying 
\begin{align*}
T(\tilde{P}) - T(P)  
&=  
 \int IF(z,\tilde{P}) \ d(\tilde{P}(z)-P(z)) \\
& \hspace{1.4cm}+ \frac{1}{2} \int\int IF_{2} (z^{(1)},z^{(2)},\tilde{P}) \ \prod_{j=1}^2 d(\tilde{P}(z^{(j)}-P(z^{(j)}))) +  R_3(\tilde{P},P),
\end{align*}
for an appropriate third-order term $R_3(\tilde{P},P)$. This has led to groundbreaking work by, for example, \citet{robins2008higher,robins2009quadratic,carone2014higher,robins2017minimax}, aimed at finding approximate higher-order influence functions that can be used for extra bias correction beyond linear/first-order corrections discussed here. There are many open problems in this domain.

\subsection{Robust statistics, and influence functions for estimators\label{sec:robust}}

IFs were first proposed to describe the stability of different estimators in cases where outliers are present, or where a portion of the sample deviates from parametric assumptions (\citealt{hampel1974influence}; see also \citealt{hampel1986robust,huber2009robust_statistics_2nd_ed_book,maronna2019robust_stat_book}). 
To see how IFs achieve these goals, consider the plug-in estimate that takes the empirical distribution of the data $\hat{P}$ as input. If we substitute $G$ with $\hat{P}$ in Definition \ref{def:if-diff-def}, the resulting Eq.~(\ref{eq:if-diff-def}) tells us how our plug-in estimate $T(\hat{P})$ would change in response to a portion of the sample ($\hat{P}$) being replaced with data from a noise distribution $Q$. Making the same substitution in Eq.~(\ref{eq:IF-single-z}), we see that the IF for $T$ also describes how the estimate $T(\hat{P})$ would change in response to an upweighting of any outlying sample point $z$. Thus, in order to produce plug-in estimates that are robust to noise contamination and outliers, a common approach is to derive functionals with \emph{bounded} IFs.

Several extensions and related uses of IFs exist for studying estimators in the form of functionals of the sample distribution. \citet{vecchia2012higher} extend IFs to describe higher order approximations of an estimator's sensitivity to sample perturbations, analogous to the approximations discussed in Section \ref{subsec:HO-IFs}. The authors also present a visual illustration of how IFs, and higher order IFs, can approximately capture robustness (see their Figure 1, which is similar to our Figure \ref{fig:single-path}). Because the $L_2$ norm used in Section \ref{subsec:Sensitivity-to-the} is relatively unaffected by the presence of outliers, an alternative choice of norm can be useful when studying robustness (see \citealt{hampel1971general}; Chapter 2 of \citealt{huber1981robust_statistics_book}; and pages 4-5 of \citealt{clarke2000review}). 
IFs can also capture the asymptotic stability of an estimator (see Chapter 5 of \citealt{van_der_Vaart2000asymptotic_statistics}).

IFs for estimators have also recently gained traction in the machine learning literature. \citet{xu2018learning} and \citet{belagiannis2015robust} use bounded loss functions when fitting a neural network, in order to reduce the influence of outliers and to improve generalization error. 
\citet{christmann2004robustness} derive conditions under which the IF for a classifier is bounded. 
\citet{koh2017understanding} compare the influence of different sample points on the predictions produced by a black box model, in order to understand what information contributed to each prediction. \citet{efron2014sd_bagging,wager2014confidence} use IFs, referred to as ``directional derivatives,'' to study the sampling variance of bagged estimators. Similarly, \citet{giordano2019swiss} propose using linear approximations of how a model will change in response to a change in the training weights, as a computationally tractable alternative to bootstrapping or cross-validation.

\subsection*{Conclusion}

For many quantitative methods, visualizations have proved to be valuable tools for communicating results and establishing intuition (e.g., for gradient descent, Lagrange multipliers, and graphical models). In this paper we provide similar tools for illustrating
IFs, based on a connection to linear approximations and Newton-Raphson. Our overall goal is to make these methods more intuitive and accessible.

The growing field of IF research shows great promise for estimating targeted quantities with higher precision, and delivering stronger scientific conclusions. Progress has been made in diverse functional estimation problems, ranging from density estimation to regression to causal inference. The approach also naturally encourages interdisciplinary collaboration, as the selection of the target parameter ($T$) benefits from deep subject area knowledge, and the initial distribution estimate ($\tilde{P}$) is often attained using powerful, flexible machine learning methods. There are many opportunities for new researchers to tackle theoretical, applied, computational, and conceptual challenges, and to push this exciting field even further.

\appendix

\section{Illustrations for the discrete case}

\begin{figure}
\begin{centering}
\includegraphics[width=1\columnwidth]{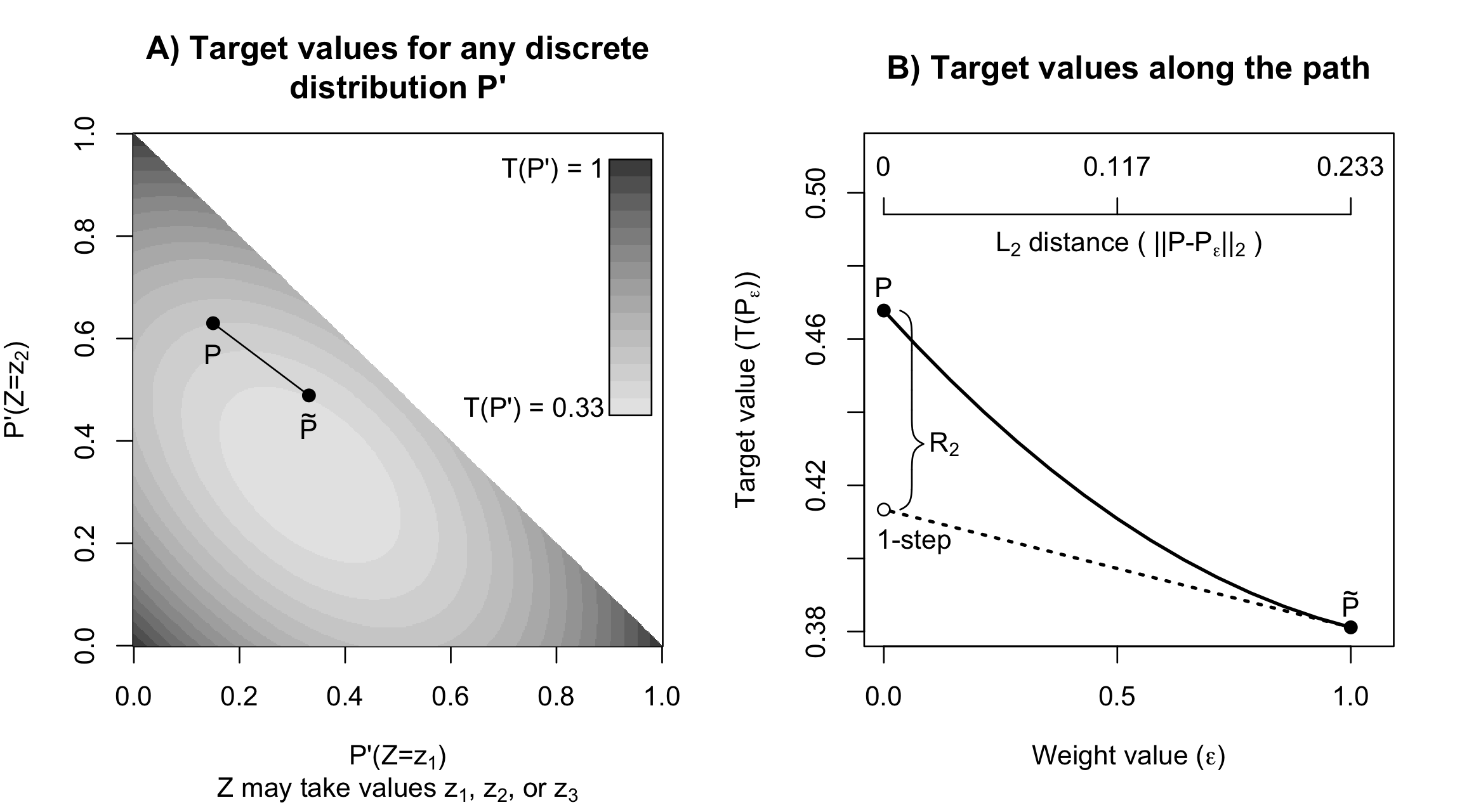}
\par\end{centering}
\caption{Linear approximation of $\mathcal{P}$ (discrete case)\label{fig:single-2d-path} -
Here we show a special case where $Z$ can take only 3 discrete values: $z_1$, $z_2$, and $z_3$. \textbf{Panel A}
shows the space of all possible distributions for $Z$, indexed (along the x and y axes) by the probability assigned to $z_1$ and $z_2$. For each possible distribution $P'$, the value of $T(P')=\sum_{i=1}^3P'(Z=z_i)^2$ is shown via shading. The upper-right triangle of the figure is left blank, as this region corresponds to invalid distributions with total mass greater than 1. Within the space of valid distributions, we show the path $\mathcal{P}$ as a straight line. As $\epsilon$ moves from 1 to 0, we move from $\tilde{P}$ to $P$ (see Eq.~(\ref{eq:P-path-def})). \textbf{Panel
B} follows the same format as Figure \ref{fig:single-path}-B. The solid line shows the target functional value $T(P_{\epsilon})$ (y-axis) as we vary $\epsilon$ (x-axis). The dotted line shows the slope of $T(P_{\epsilon})$
with respect to $\epsilon$ at $\epsilon=1$. As in Figure \ref{fig:single-path}-B, we show the distributional distance on a secondary horizontal axis at the top
of the figure. In this case though, distributional distance $||P-P_{\epsilon}||_{2} = \sqrt{\sum_{i=1}^3\{P(Z=z_i)-P_{\epsilon}(Z=z_i)\}^2}$  can also be visually approximated by Euclidean distance in Panel A (ignoring the third summation term $\{P(Z=z_3)-P_{\epsilon}(Z=z_3)\}^2$).
}
\end{figure}

\begin{figure}
\begin{centering}
\includegraphics[width=1\columnwidth]{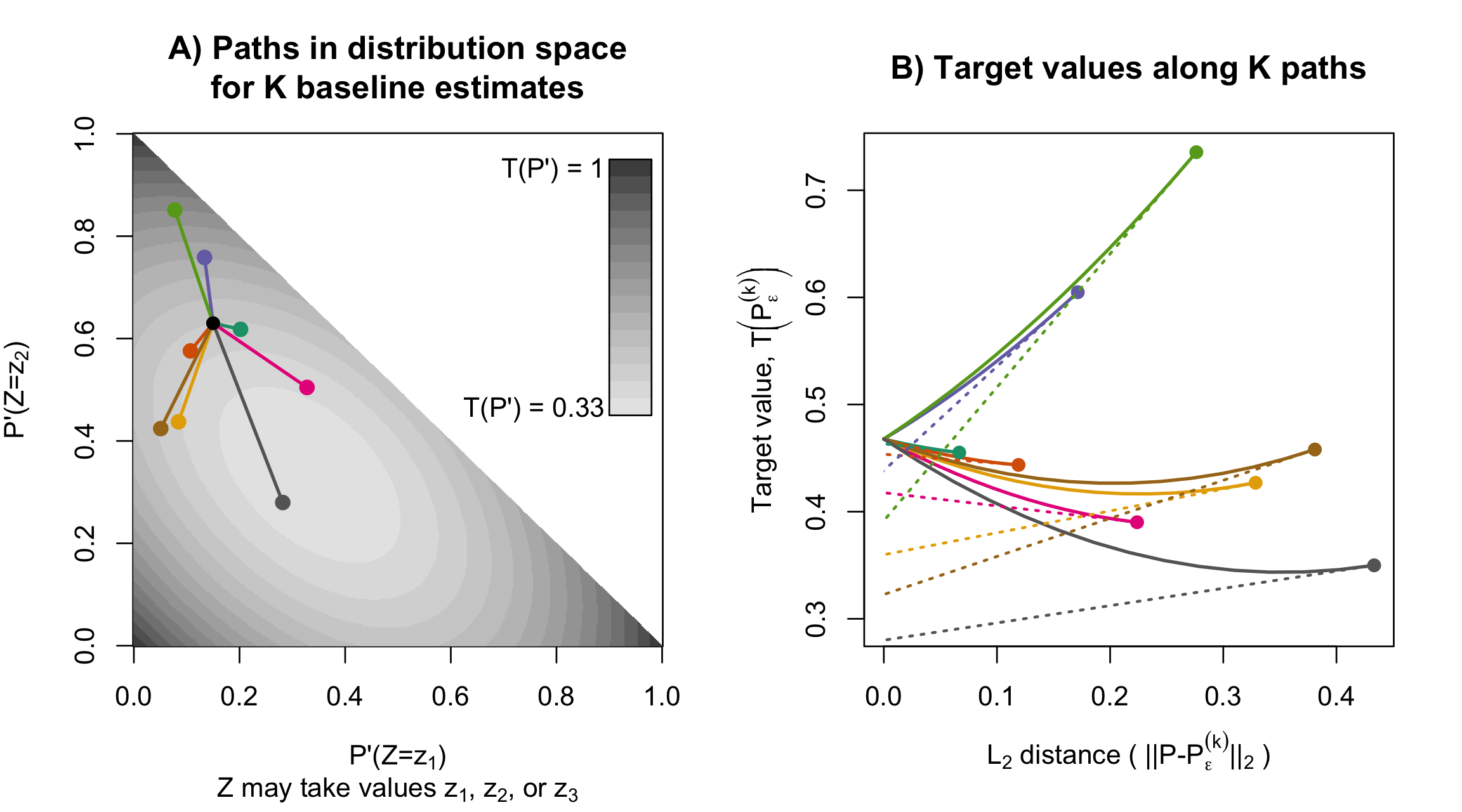}
\par\end{centering}
\caption{Linear approximations overlaid for several paths (discrete case) \label{fig:several-3d-paths}
- Above, we overlay the same illustrations as in Figure \ref{fig:single-2d-path},
but for several alternative initial distribution estimates $\tilde{P}^{(1)},\dots,\tilde{P}^{(K)}$. The result is analogous to Figure \ref{fig:several-paths}, for the special case where $Z$ is discrete.
Here, \textbf{Panel A} shows several paths through the space of distributions, each defined in the same way as in Eq.~(\ref{eq:P-path-def}), but starting from a different initial estimate $\tilde{P}^{(k)}$. \textbf{Panel B} shows the values
of the target parameter at each point $\tilde{P}_{\epsilon}^{(k)}$
along each path $\mathcal{P}^{(k)}$, as well as a linear approximation
of each path. The x-axis shows distributional distance from $P$, the dotted lines show linear approximations, and the y-intercepts of each dotted line correspond to different 1-step estimators. Again, we see that the accuracy of each estimator will depend on the distance $||P-\tilde{P}^{(k)}||_{2}$. This distance can also be approximated from Euclidean distance in Panel A.}
\end{figure}

Figures \ref{fig:single-2d-path} and \ref{fig:several-3d-paths} show  alternate versions of Figures \ref{fig:single-path} and \ref{fig:several-paths} for the special case where $Z$ can take only 3 discrete values: $z_1$, $z_2$, and $z_3$. In this case, any probability distribution for $Z$ can be fully described by the probability it assigns to $z_1$ and $z_2$. This simplicity allows us to depict the full space of possible distributions, and the value of $T$ for each distribution, in a 2-dimensional figure.

Note that Figures \ref{fig:single-2d-path}-B and \ref{fig:several-3d-paths}-B are essentially unchanged from  Figures \ref{fig:single-path}-B and \ref{fig:several-paths}-B. This is because it is always possible to visualize 1-dimensional paths through the space of possible distributions, regardless of the dimensionality of that space. In other words, we can visualize paths through the space of distributions regardless of whether we can visualize the space itself (as in Figures \ref{fig:single-2d-path} \& \ref{fig:several-3d-paths}).

\section{Score-based definition of the IF}

An alternative definition of the IF describes derivatives along paths
not necessarily of the form $G+\epsilon(Q-G)$. This can be especially beneficial when prior knowledge restricts the space distributions that we consider possible, and when this allowed distribution space is not closed under convex combinations of the form $G+\epsilon(Q-G)$ (see Section \ref{sec:semipara}). We can define a more
general pathway as simply the set of distributions consistent with
a certain likelihood model $\mathcal{L}(z;e)$, with scalar parameter
$e\in[0,1]$. Let $w_{e}(z)$ be the density associated with the likelihood
function $\mathcal{L}(z;e)$, and let $W_{e}$ be the associated distribution
function. With this notation, we can now give an alternate definition for the IF (see \citealt{bickel1993efficient,tsiatis2006semiparametric}).
\begin{definition}
\label{def:For-any-likelihood}(``score-based'' IF) The influence
function for $T$ is the function $IF$ satisfying 
\begin{equation}
\left.\frac{\partial T(W_{e})}{\partial e}\right|_{e=0}=E_{W_{0}}[ IF(Z,W_{0})s_{0}(Z)],\label{eq:any-likelihood}
\end{equation}
and $E_{W_{0}}IF(Z,W_{0})=0$ for any likelihood $W_{e}$, where $s_{e}$ is the score function
$s_{e}(z)=\frac{\partial}{\partial e}\log w_{e}(z)$, with $w_{e}$
being the density of $W_{e}$.
\end{definition}

It is fairly straightforward to show that Definition \ref{def:For-any-likelihood} implies Definition \ref{def:if-diff-def}. That is, if a function satisfies Definition \ref{def:For-any-likelihood}, it must also satisfy Definition \ref{def:if-diff-def} (in the case of no prior restrictions on the space of allowed distributions). To see this, note that for any two distributions $G$ and $Q$ we can define a likelihood $W_e:=G+e(Q-G)$ with score function
\begin{align*}
s_{0}(z) & =\left.\frac{\partial}{\partial e}\log\left[g(z)+e\left\{ q(z)-g(z)\right\} \right]\right|_{e=0}\\
 & =\frac{q(z)-g(z).}{g(z)}.
\end{align*}

Definition \ref{def:For-any-likelihood} now implies that 
\begin{align*}
\left.\frac{\partial T(W_{e})}{\partial e}\right|_{e=0} & =\int IF(z,W_{0})s_{0}(z)q_{0}(z)dz\\
 & =\int IF(z,G)\left\{ \frac{q(z)-g(z)}{g(z)}\right\} g(z)dz\\
 & =\int IF(z,G)\left\{ q(z)-g(z)\right\} dz,
\end{align*}
which shows that $IF$ satisfies Definition \ref{def:if-diff-def}.

\section{Derivation of IF and $R_2$ term for the squared integrated density functional}

Let $G$ and $Q$ be defined as in Definition \ref{def:if-diff-def}, with densities
$g$ and $q$ that are dominated by an integrable function $\nu$. For $T(G)=\int g(z)^{2}dz$, the influence function is
equal to $IF(z,G)=2(g(z)-T(G))$ \citep{bickel1988estimating,robins2008higher}. To see this, we first show Eq.~(\ref{eq:if-diff-def}).
\begin{align*}
 & \left.\frac{\partial T(G+\epsilon(Q-G))}{\partial\epsilon}\right|_{\epsilon=0}\\
 & \hspace{1cm}=\left.\frac{\partial}{\partial e}\int[g(z)+\epsilon\{q(z)-g(z)\}]^{2}dz\right|_{e=0}\\
 & \hspace{1cm}\left.\int\frac{\partial}{\partial e}[g(z)+\epsilon\{q(z)-g(z)\}]^{2}dz\right|_{e=0} &  & \text{Dominated Convergence Thm}\\
 & \hspace{1cm}\left.\int2[g(z)+\epsilon\{q(z)-g(z)\}][q(z)-g(z)]dz\right|_{e=0}\\
 & \hspace{1cm}\int2[g(z)-T(G)][q(z)-g(z)]dz &  & \text{from }\int T(G)[q(z)-g(z)]dz=0\\
 & \hspace{1cm}\int IF(z,G)[q(z)-g(z)]dz.
\end{align*}

This, in combination with the fact that
\begin{equation*}\int IF(z,G)g(z)dz=2\int \{g(z)^2 - T(G)g(z) \}dz=0,
\end{equation*} establishes that $IF(z,G)=2(g(z)-T(G))$ is the influence function for $T(G)=\int g(z)^{2}dz$.

Given a fixed distribution estimate $\tilde{P}$, the bias ($R_{2}$ term) of $\hat{T}_{\text{1-step}}$ is equal to 
\begin{align*}
E_P(\hat{T}_{\text{1-step}})-T(P) & =\left\{ T(\tilde{P})+\int IF(z,\tilde{P})p(z)dz\right\} -T(P)\\
 & =T(\tilde{P})+
 \int 2 \tilde{p}(z)p(z)dz-2T(\tilde{P})
 -T(P)\\
 & =-T(\tilde{P})+
 \int 2 \tilde{p}(z)p(z)dz
 -T(P)\\
 & =-\int\{\tilde{p}(z)-p(z)\}^{2}dz.
\end{align*}

\section{Showing distance results for $P_\epsilon$ and $P_{\Delta}^\text{rescaled}$\label{sec:dist-results}}

To show $||P-P_{\epsilon}||_{2}/||P-\tilde{P}||_{2}=\epsilon$, we have
\begin{align}
||P-P_{\epsilon}||_{2} & =\sqrt{\int[p(z)-p_{\epsilon}(z)]^{2}dz}\nonumber \\
 & =\sqrt{\int[p(z)-(1-\epsilon)p(z)-\epsilon\tilde{p}(z)]^{2}dz}\nonumber \\
 & =\sqrt{\int\epsilon^{2}[p(z)-\tilde{p}(z)]^{2}dz}\nonumber \\
 & =\epsilon||P-\tilde{P}||_{2}.\label{eq:dist-equivalence}
\end{align}

The fact that $||P_{\Delta}^\text{rescaled}-P||_2=\Delta$ now follows from
\begin{equation*}
 ||P-P_{\Delta}^\text{rescaled}||_2 
=||P-P_{\Delta/||P-\tilde{P}||}||_2
=\frac{\Delta||P-\tilde{P}||}{||P-\tilde{P}||}=\Delta,
\end{equation*}
where the first equality follows from the definition of $P_{\Delta}^\text{rescaled}$, and the second equality comes from Eq.~(\ref{eq:dist-equivalence}).

\section{Proof of Remark \ref{rem:R1R2}}

We begin with Eq.~(\ref{eq:remark-R1}), which we will show using Taylor's Theorem and Condition \ref{cond:no-squiggles} for $j=1$. Taylor's Theorem implies that there exists
a value $\bar{\epsilon}\in[0,1]$ such that 
\begin{equation}
T(P_{1})-T(P_{0})=\left.\frac{\partial}{\partial\epsilon}T(P_{\epsilon})\right|_{\epsilon=\bar{\epsilon}}.\label{eq:Tay-1-proof-rem}
\end{equation}
In order to study the right-hand side, we introduce a function to help map between distributions in the form of $P_\epsilon$ and $P_{\Delta}^\text{rescaled}$. Let $D(\epsilon):=\epsilon||\tilde{P}-P||_{2}$,
with inverse function $D^{-1}(\Delta):=\Delta /||\tilde{P}-P||_{2}$, 
such that $P_{\epsilon}=P_{D(\epsilon)}^{\text{rescaled}}$
and $P_{\Delta}^{\text{rescaled}}=P_{D^{-1}(\Delta)}$. (For notational convenience, we omit the dependence on $\tilde{P}$ when writing
$D$, $D^{-1}$, $P_{\Delta}^{\text{rescaled}}$, and $\bar{\epsilon}$.) Returning
to Eq.~(\ref{eq:Tay-1-proof-rem}), we have
\begin{align}
\frac{\partial T(P_{\epsilon})}{\partial\epsilon}=\frac{\partial T(P_{D(\epsilon)}^{\text{rescaled}})}{\partial\epsilon} & =\left\{ \frac{\partial T(P_{D(\epsilon)}^{\text{rescaled}})}{\partial D(\epsilon)}\right\} \left\{ \frac{\partial D(\epsilon)}{\partial\epsilon}\right\}  &  & \text{by the chain rule}\nonumber \\
 & =\left\{ \frac{\partial T(P_{D(\epsilon)}^{\text{rescaled}})}{\partial D(\epsilon)}\right\} ||\tilde{P}-P||_{2}.\label{eq:base-case}
\end{align}
Plugging this into Eq~(\ref{eq:Tay-1-proof-rem}), we have
\begin{align}
T(\tilde{P})-T(P) & =\left.\frac{\partial T(P_{D(\epsilon)}^{\text{rescaled}})}{\partial D(\epsilon)}\right|_{\epsilon=\bar{\epsilon}}||\tilde{P}-P||_{2}\nonumber \\
 & =\left.\frac{\partial T(P_{\Delta}^{\text{rescaled}})}{\partial\Delta}\right|_{\Delta=D(\bar{\epsilon})}||\tilde{P}-P||_{2}\nonumber \\
 & =O(1)\times||\tilde{P}-P||_{2}\label{eq:lim1}\\
 & =O(||\tilde{P}-P||_{2}),\label{eq:lim2}
\end{align}
Where the limits in Eq.~(\ref{eq:lim1}) \& Eq.~(\ref{eq:lim2}) are
taken as $||\tilde{P}-P||_{2}\rightarrow0$. To arrive at Eq.~(\ref{eq:lim1}), note that when $||\tilde{P}-P||_{2}\rightarrow0$ we have $D(\epsilon)=\bar{\epsilon}||\tilde{P}-P||_{2}\leq||\tilde{P}-P||_{2}\rightarrow0$, and therefore $\left.\frac{\partial T(P_{\Delta}^{\text{rescaled}})}{\partial\Delta}\right|_{\Delta=D(\bar{\epsilon})}=O(1)$
by Condition \ref{cond:no-squiggles} (with $j=1$).

Turning to Eq.~(\ref{eq:remark-R2}), the first equality of  follows from Eq.~(\ref{eq:1dim-taylor-eq}) and Eq.~(\ref{eq:IF-approx-sample}).

We can show the second equality of Eq.~(\ref{eq:remark-R2}) by again applying Taylor's Theorem and Condition \ref{cond:no-squiggles}, this time with $j=2$. Taylor's Theorem implies that there exists a value
$\bar{\epsilon}\in[0,1]$ satisfying $R_{2}=(-1/2)\left.\frac{\partial^{2}}{\partial\epsilon^{2}}T(P_{\epsilon})\right|_{\epsilon=\bar{\epsilon}}$, as discussed in the text following Eq.~(\ref{eq:1dim-taylor-eq}).
To study this second derivative of $T(P_{\epsilon})$, we will show
that, for finite $j$,
\begin{equation}
\frac{\partial^{j}T(P_{\epsilon})}{\partial\epsilon^{j}}
=\left\{ \frac{\partial^{j}T(P_{D(\epsilon)}^{\text{rescaled}})}{\partial D(\epsilon)^{j}}\right\} ||\tilde{P}-P||_{2}^{j}.\label{eq:induction}
\end{equation}
The proof of Eq.~(\ref{eq:induction}) is by induction. We have already
shown the base case of $j=1$ in Eq.~(\ref{eq:base-case}). For the
induction step, given that Eq.~(\ref{eq:induction}) holds for $j-1$,
we can show that Eq.~(\ref{eq:induction}) holds for $j$ as follows.
\begin{align*}
\frac{\partial^{j}T(P_{\epsilon})}{\partial\epsilon^{j}} 
& =\frac{\partial}{\partial\epsilon}\left\{ \frac{\partial^{j-1}T(P_{\epsilon})}{\partial\epsilon^{j-1}}\right\} \\
 & =\left[\frac{\partial}{\partial \epsilon}\left\{ \frac{\partial^{j-1} T(P_{D(\epsilon)}^{\text{rescaled}})}{\partial D(\epsilon)^{j-1}}||\tilde{P}-P||_{2}^{j-1}\right\} \right] &  & \text{by Eq.~(\ref{eq:base-case}) for }j-1\\
 & =\left[\frac{\partial}{\partial D(\epsilon)}\left\{ \frac{\partial^{j-1} T(P_{D(\epsilon)}^{\text{rescaled}})}{\partial D(\epsilon)^{j-1}}||\tilde{P}-P||_{2}^{j-1}\right\} \right]\left[\frac{\partial D(\epsilon)}{\partial\epsilon}\right] &  & \text{by the chain rule}\\
 & =\frac{\partial^{j}T(P_{D(\epsilon)}^{\text{rescaled}})}{\partial D(\epsilon)^{j}}||\tilde{P}-P||_{2}^{j}.
\end{align*}
Finally, applying Eq.~(\ref{eq:induction}), we have
\begin{align}
R_{2}& =\frac{-1}{2}\left.\frac{\partial^{2}}{\partial\epsilon^{2}}T(P_{\epsilon})\right|_{\epsilon=\bar{\epsilon}} & & \nonumber \\
& =\frac{-1}{2}\left.\left\{ \frac{\partial^{2}T(P_{D(\epsilon)}^{\text{rescaled}})}{\partial D(\epsilon)^{2}}\right\} ||\tilde{P}-P||_{2}^{2}\right|_{\epsilon=\bar{\epsilon}}&& \nonumber \\
 & =\frac{-1}{2}\left.\left\{ \frac{\partial^{2}T(P_{\Delta}^{\text{rescaled}})}{\partial\Delta^{2}}\right\} \right|_{\Delta=D(\bar{\epsilon})}||\tilde{P}-P||_{2}^{2}&& \nonumber \\
 & =O(||\tilde{P}-P||_{2}^{2}).\label{eq:lim3}
\end{align}

As in Eq.~(\ref{eq:lim1}), the limit in Eq.~(\ref{eq:lim3}) is
taken as $||\tilde{P}-P||_{2}\rightarrow0$. Eq.~(\ref{eq:lim3}) comes from the fact that when $||\tilde{P}-P||_{2}\rightarrow0$ we have $D(\epsilon)=\bar{\epsilon}||\tilde{P}-P||_{2}\leq||\tilde{P}-P||_{2}\rightarrow0$, and therefore $\left.\frac{\partial^2 T(P_{\Delta}^{\text{rescaled}})}{\partial\Delta^2}\right|_{\Delta=D(\bar{\epsilon})}=O(1)$
by Condition \ref{cond:no-squiggles} (with $j=2$).

\bibliographystyle{apalike}
\bibliography{IF-pic}

\end{document}